\documentclass[a4paper,12pt]{article}

\catcode`\@=11 \long\def\@makefntext#1{ \protect\noindent \hbox to 3.2pt
{\hskip-.9pt
$^{{\eightrm\@thefnmark}}$\hfil}#1\hfill}       

\def\@makefnmark{\hbox to 0pt{$^{\@thefnmark}$\hss}}    

\def\ps@myheadings{\let\@mkboth\@gobbletwo
\def\@oddhead{\hbox{}
\rightmark\hfil\eightrm\thepage}
\def\@oddfoot{}\def\@evenhead{\eightrm\thepage\hfil
\leftmark\hbox{}}\def\@evenfoot{}
\def\sectionmark##1{}\def\subsectionmark##1{}}

\oddsidemargin=\evensidemargin 
\newcounter{sectionc}\newcounter{subsectionc}\newcounter{subsubsectionc}
\renewcommand{\section}[1] {\vspace{12pt}\addtocounter{sectionc}{1}
\setcounter{subsectionc}{0}\setcounter{subsubsectionc}{0}\noindent
    {\tenbf\thesectionc. #1}\par\vspace{5pt}}
\renewcommand{\subsection}[1] {\vspace{12pt}\addtocounter{subsectionc}{1}
    \setcounter{subsubsectionc}{0}\noindent
    {\bf\thesectionc.\thesubsectionc. {\kern1pt \bfit #1}}\par\vspace{5pt}}
\renewcommand{\subsubsection}[1] {\vspace{12pt}\addtocounter{subsubsectionc}{1}
    \noindent{\tenrm\thesectionc.\thesubsectionc.\thesubsubsectionc.
    {\kern1pt \tenit #1}}\par\vspace{5pt}}

\newcounter{appendixc}
\newcounter{subappendixc}[appendixc]
\newcounter{subsubappendixc}[subappendixc]
\renewcommand{\thesubappendixc}{\Alph{appendixc}.\arabic{subappendixc}}
\renewcommand{\thesubsubappendixc}
    {\Alph{appendixc}.\arabic{subappendixc}.\arabic{subsubappendixc}}

\renewcommand{\appendix}[1] {\vspace{12pt}
        \refstepcounter{appendixc}
        \setcounter{figure}{0}
        \setcounter{table}{0}
        \setcounter{lemma}{0}
        \setcounter{theorem}{0}
        \setcounter{corollary}{0}
        \setcounter{definition}{0}
        \setcounter{equation}{0}
        \renewcommand{\thefigure}{\Alph{appendixc}.\arabic{figure}}
        \renewcommand{\thetable}{\Alph{appendixc}.\arabic{table}}
        \renewcommand{\theappendixc}{\Alph{appendixc}}
        \renewcommand{\thelemma}{\Alph{appendixc}.\arabic{lemma}}
        \renewcommand{\thetheorem}{\Alph{appendixc}.\arabic{theorem}}
        \renewcommand{\thedefinition}{\Alph{appendixc}.\arabic{definition}}
        \renewcommand{\thecorollary}{\Alph{appendixc}.\arabic{corollary}}
        \renewcommand{\theequation}{\Alph{appendixc}.\arabic{equation}}
        \noindent{\tenbf Appendix \theappendixc #1}\par\vspace{5pt}}
\newcommand{\subappendix}[1] {\vspace{12pt}
        \refstepcounter{subappendixc}
        \noindent{\bf Appendix \thesubappendixc. {\kern1pt \bfit #1}}
    \par\vspace{5pt}}
\newcommand{\subsubappendix}[1] {\vspace{12pt}
        \refstepcounter{subsubappendixc}
        \noindent{\rm Appendix \thesubsubappendixc. {\kern1pt \tenit #1}}
    \par\vspace{5pt}}

\newcommand{\textlineskip}{\baselineskip=13pt}
\newcommand{\smalllineskip}{\baselineskip=10pt}

\def\eightcirc{
\begin{picture}(0,0)
\put(4.4,1.8){\circle{6.5}}
\end{picture}}
\def\eightcopyright{\eightcirc\kern2.7pt\hbox{\eightrm c}}

\def\abstracts#1#2#3{{
    \centering{\begin{minipage}{4.5in}\footnotesize\baselineskip=10pt
    \parindent=0pt #1\par
    \parindent=15pt #2\par
    \parindent=15pt #3
    \end{minipage}}\par}}

\newcounter{itemlistc}
\newcounter{romanlistc}
\newcounter{alphlistc}
\newcounter{arabiclistc}

\newcommand{\fcaption}[1]{
        \refstepcounter{figure}
        \setbox\@tempboxa = \hbox{\footnotesize Fig.~\thefigure. #1}
        \ifdim \wd\@tempboxa > 5in
           {\begin{center}
        \parbox{5in}{\footnotesize\smalllineskip Fig.~\thefigure. #1}
            \end{center}}
        \else
             {\begin{center}
             {\footnotesize Fig.~\thefigure. #1}
              \end{center}}
        \fi}

\newcommand{\tcaption}[1]{
        \refstepcounter{table}
        \setbox\@tempboxa = \hbox{\footnotesize Table~\thetable. #1}
        \ifdim \wd\@tempboxa > 5in
           {\begin{center}
        \parbox{5in}{\footnotesize\smalllineskip Table~\thetable. #1}
            \end{center}}
        \else
             {\begin{center}
             {\footnotesize Table~\thetable. #1}
              \end{center}}
        \fi}

\def\@citex[#1]#2{\if@filesw\immediate\write\@auxout
    {\string\citation{#2}}\fi
\def\@citea{}\@cite{\@for\@citeb:=#2\do
    {\@citea\def\@citea{,}\@ifundefined
    {b@\@citeb}{{\bf ?}\@warning
    {Citation `\@citeb' on page \thepage \space undefined}}
    {\csname b@\@citeb\endcsname}}}{#1}}

\newif\if@cghi
\def\cite{\@cghitrue\@ifnextchar [{\@tempswatrue
    \@citex}{\@tempswafalse\@citex[]}}
\def\citelow{\@cghifalse\@ifnextchar [{\@tempswatrue
    \@citex}{\@tempswafalse\@citex[]}}
\def\@cite#1#2{{$\null^{#1}$\if@tempswa\typeout
    {IJCGA warning: optional citation argument
    ignored: `#2'} \fi}}

\def\pmb#1{\setbox0=\hbox{#1}
    \kern-.025em\copy0\kern-\wd0
    \kern.05em\copy0\kern-\wd0
    \kern-.025em\raise.0433em\box0}

\def\fnt#1#2{\footnotetext{\kern-.3em
    {$^{\mbox{\scriptsize #1}}$}{#2}}}

\def\fpage#1{\begingroup
\voffset=.3in \thispagestyle{empty}\begin{table}[b]\centerline{\footnotesize #1}
    \end{table}\endgroup}

\def\runninghead#1#2{\pagestyle{myheadings}
\markboth{{\protect\footnotesize\it{\quad #1}}\hfill}
{\hfill{\protect\footnotesize\it{#2\quad}}}} \headsep=15pt

\font\tenrm=cmr10 \font\tenit=cmti10 \font\tenbf=cmbx10 \font\bfit=cmbxti10 at 10pt
\font\ninerm=cmr9   \font\eightrm=cmr8

\def\qed{\hbox{${\vcenter{\vbox{            
   \hrule height 0.4pt\hbox{\vrule width 0.4pt height 6pt
   \kern5pt\vrule width 0.4pt}\hrule height 0.4pt}}}$}}

\begin{document}
\setlength{\textheight}{7.7truein}  

\runninghead{$\theta$-term and cosmological constant from CJD action}{$\theta$-term
and cosmological constant from CJD action}

\normalsize\textlineskip \thispagestyle{empty} \setcounter{page}{1}



\fpage{1} \centerline{\bf $\theta$-TERM AND COSMOLOGICAL CONSTANT FROM CJD ACTION}
\centerline{\footnotesize PRASANTA MAHATO\footnote{e-mail:
pmahato@dataone.in}}
\centerline{\footnotesize\it Department of Mathematics, Narasinha
Dutt College} \baselineskip=10pt \centerline{\footnotesize\it HOWRAH,West
Bengal,INDIA 711101} \vspace*{10pt}

\vspace*{0.225truein}




 \abstracts{ In the gravity without metric formalism of
Capovilla, Jacobson and Dell, the topological $\theta$-term appears through a
canonical transformation.The origin of this canonical  transformation is probed
here. It is shown here that when $\theta$-term appears cosmological $\lambda$-term
also appears simultaneously.}{PACS numbers : 04.20.Cv, 04.20.Fy}{Key words :
Canonical transformation, Torsion, $\theta$-term, Cosmological constant}



\vspace*{-0.5pt} \vspace*{0.21truein}
 Einstein originally formulated the theory of gravitation as a set of differential
    equations obeyed by the metric tensor of space-time. Ashtekar\cite{Ash88,Ash87} has rewritten Einstein's
    theory, in its hamiltonian formulation as a set of differential equations obeyed by an
    SO(3) connection and its canonically conjugate momentum. A Lagrangian formulation, in which
    the variables are the space-time tetrads and the self-dual spin connection, was given by Samuel\cite{Sam87},
    Jacobson and Smolin\cite{Smo88}.

  If tetrad is invertible, one may eliminate the spin connection to obtain
    the conventional Hilbert action, plus a (complex) surface term, so that the Lagrangian
    formulation is given entirely in terms of the tetrad. However, it is clearly natural to
    enquire whether one can give a Lagrangian formulation of Ashtekar's theory in which the metric or tetrad,
     has been completely eliminated in favour of the connection. This has been done by Capovilla, Jacobson and
      Dell\cite{Del89,Del91}.


 Bengtsson and Peldan\cite{Ben90} have shown that if one performs a particular canonical
transformation involving Ashtekar variables and corresponding SO(3) gauge fields,
the expression for the Hamiltonian constraint changes when other constraints remain
unaffected. This corresponds precisely to the addition of a \textquotedblleft
CP-violating\textquotedblright \hspace{1 mm}  $\theta$-term to the CJD Lagrangian.
Mullick and Bandyopadhyay\cite{Ban95} have shown that this \textquotedblleft
CP-violating\textquotedblright \hspace{1 mm}  $\theta$-term is responsible for
non-zero torsion.  This $\theta$-term effectively corresponds to the chiral anomaly
when a fermion chiral current interacts with a gauge field.
     In a recent paper\cite{Ban00}, it has been shown that chiral anomaly gives rise
      to the mass of a fermion which implies that for a massive fermion the divergence of the
      axial vector current
      (which is associated with torsion via $\theta$-term) is non-vanishing.

 The well-known action in terms of Ashtekar variables in its (3+1)-dimensional form can be written
      directly as\cite{Ben90,Ben92,Pel93},
     \begin{eqnarray}
     S &=& \int\dot{A}_{ai}E_i^a-NH-N^aH_a-\lambda_iG_i\\
     \mbox{where}\hspace{2mm} H& =& \frac{1}{2}if_{ijk}E_i^aE_j^bF_{abk}\\
             & =&  \frac{1}{2}i\epsilon_{abc}f_{ijk}E_i^aE_j^bB_k^c\\
             H_a& =& E_i^bF_{abi} = \epsilon_{abc}E_i^bB_i^c\\
      G_i& =& D_aE_i^a = \partial _a E_i^a+if_{ijk}A_{aj}E_k^a .
      \end{eqnarray}
      \\Here $a,b,c$ are spatial indices, $i,j,k$ are SO(3) indices, $F_{abi}$ is
      an SO(3) curvature and $B_i^a = \frac{1}{2}\epsilon^{abc}F_{bci}$ is
      the corresponding \textquotedblleft magnetic\textquotedblright \hspace{1 mm} field.
       N, $N^a$ are the lapse and shift functions respectively and $\lambda_i$ is a multiplier. These yield three kinds of constraints, viz.,
      the Hamiltonian   constraint H$\approx 0$, the vector constraint H$_a\approx 0$ and the SO(3) vector's worth
      G$_i\approx 0$  of \textquotedblleft internal\textquotedblright\hspace{1 mm} constraints which here takes the form of Gauss' law for the
      \textquotedblleft electric\textquotedblright \hspace{1 mm}
       field E$_i^a$. The tensor q$^{ab} = gg^{ab},$ where $g^{ab}$ is the metric tensor
       on the foliating hypersurface and $g$ is its determinant, is given by $q^{ab} =
       E_i^aE_i^b.$ Here the fundamental Poisson bracket is \hspace{2 mm}
       $\{A_{ai}(x),E^{bj}(y)\}=\delta_a^b\delta_i^j\delta^3(x-y).$

        Capovilla, Jacobson and Dell\cite{Del89,Del91}(CJD)\hspace{1 mm} found a very elegant
         Lagrangian   formulation of the above. The CJD action can be written as\cite{Ben90,Ben92},

       \begin{eqnarray}S^{CJD}& =& \frac{1}{8}\int\eta(\Omega_{ij}\Omega_{ij}+a\Omega_{ii}\Omega_{jj})\\
        \mbox{where} \hspace{2mm}\Omega_{ij}&=&\epsilon^{\alpha \beta \gamma \delta}F_{\alpha \beta i}F_{\gamma \delta j},
       \end{eqnarray} here $\alpha,\beta,\gamma,\delta$ are space-time indices,
       the Lagrangian multiplier $\eta$ is a scalar density of weight -1, such that $\frac{1}{\eta}d^4x $ is an invariant volume element,  and F$_{\alpha\beta i}$ is an SO(3)
        field strength. The 3+1 dimensional decomposition of this action yields Ashtekar's action
         directly provided that the parameter $a = -1/2$ and that the determinant of the
         \textquotedblleft magnetic\textquotedblright \hspace{1 mm} field
         $B_i^a$ is nonzero.

 In this way, the equivalence to Einstein's theory is established.
        Indeed, with $a=-\frac{1}{2}$ the equivalence to Einstein's theory may also
        be shown directly in a manifestly covariant way. This demonstration hinges
         on the tetrad formalism and the space-time metric may be given directly
          in terms of the curvature  when we can write\cite{Ben92,Pel90},

          \begin{eqnarray}(-g)^\frac{1}{2}g^{\alpha\beta}& = &-\frac{2i}{3}\eta K^{\alpha\beta}\\&
          = &-\frac{2i}{3}\eta f_{ijk}\epsilon^{\alpha\gamma\delta\rho}\epsilon^{\beta\mu\nu
          \sigma} F_{\gamma\delta i}F_{\rho\sigma j}F_{\mu\nu k}.\end{eqnarray}

           The constraint that is obtained when the CJD action is varied with
          respect to the Lagrangian multiplier $\eta$ is actually the Hamiltonian
          constraint in disguise\cite{Ben90}:

         \begin{eqnarray}  \psi& = &\Omega_{ij}\Omega_{ij}-\frac{1}{2}\Omega_{ii}\Omega_{jj}\\& = &i(2\eta^2detB)^{-1}H.\end {eqnarray}

         In this formalism the momentum canonically conjugate to the connection takes the form\cite{Ben90}
          \begin{eqnarray}E_i^a &=& \psi_{ij}B_j^a\\
          \mbox{where}\hspace{2 mm}\psi_{ij} &=& 2\eta (\Omega_{ij} -
          \frac{1}{2}\delta_{ij}\Omega_{kk}).\end{eqnarray}

         Here we note that such a matrix $\psi_{ij}$ always exists provided that the magnetic
         field is non-degenerate. If we insert this  matrix in the vector constraint
         $H_a \approx 0$, we find that the vanishing of the vector constraint is equivalent to
          the statement that the matrix is symmetric. We also note that as long
          as \textsl{detB} $\ne0,\psi \approx 0,H \approx 0$ are equivalent statements. Gauss' law
          $G_i \approx 0$ follows when the action is varied with respect to $A_{0i}$.

 Bengtsson and Peldan\cite{Ben90} have shown that if we perform the canonical transformation,
            \begin{eqnarray}A_{ai}& \longrightarrow &A_{ai},\\ E_i^a
           &\longrightarrow &E_i^a-\theta B_i^a,\end{eqnarray}
           the expression for the Hamiltonian constraint changes though the remaining
           constraints are unaffected. This corresponds precisely to the addition of a \textquotedblleft CP-violating''
           $\theta$-term to the CJD lagrangian, when the new action is given by,

         \begin{eqnarray}  S&=&\frac{1}{8}\int\{\theta\Omega_{ii}+\eta(\Omega_{ij}\Omega_{ij}-\frac{1}{2}\Omega_{ii}\Omega_{jj})\}.\end{eqnarray}

          In CJD formalism $A_{ai}$ and $E_i^a$ are Ashtekar variables whereas
          $A_{0i}$ acts as a Lagrange's multiplier. Variation of  $A_{0i}$ yields the
          Gauss' law $G_i\approx0$. In the above canonical transformation we see that
          as $A_{ai} \longrightarrow A_{ai}$ we have $B_i^a \longrightarrow B_i^a$.
          Then $\psi_{ij} = E_i^a(B^{-1})_{aj}$ transforms to $\psi_{ij} -
          \theta\delta_{ij}$. This gives $\Omega_{ij} =
          \frac{1}{2\eta}(\psi_{ij}-\delta_{ij}\psi_{kk})\rightarrow
          \Omega_{ij}+\frac{\theta}{\eta}\delta_{ij}$. Now $\Omega_{ij}= \epsilon^{\alpha \beta \gamma \delta}F_{\alpha \beta i}F_{\gamma \delta
          j} = 4(F_{0ai}B_j^a +F_{0aj}B_i^a)$ implies $F_{0ai}$ must transform accordingly to
          produce the desired canonical transformation. We can write $F_{0ai}=
          D_aA_{0i} - \dot{A}_{ai}$. Therefore if we consider $A_{0i}\longrightarrow
          A_{0i}+\epsilon_i$, where $\epsilon_i$ is a vector having only internal index, then we get
          \begin{eqnarray}D_a\epsilon_iB_j^a + D_a\epsilon_jB_i^a &=&
          \frac{\theta}{4\eta}\delta_{ij}.\end{eqnarray}
          This equation has a solution for $D_{a}\epsilon_i$ given by \begin{eqnarray}D_a\epsilon_i
          &=& \frac{\theta}{8\eta}(B^{-1})_{ai}.\end{eqnarray}
          Thus to incorporate a \textquotedblleft CP-violating" $\theta$ term in the
          CJD action by means of a canonical transformation one should transform the
          gauge fields in the following way

          \begin{eqnarray} A_{ai}&\longrightarrow& A_{ai},\\ A_{0i}&\longrightarrow& A_{0i}+\epsilon_i,\end{eqnarray}
           where $\epsilon_i$ is an internal vector satisfying equation(18), then we see that,

             \begin{eqnarray}  \Omega_{ij} &\longrightarrow &\Omega_{ij} + \frac{\theta}{\eta}\delta_{ij},
             \\\psi_{ij}&\longrightarrow&\psi_{ij}-\theta\delta_{ij},\\ E_i^a&\longrightarrow& E_i^a-\theta B_i^a,\\
            \mbox{and}\hspace{2 mm} B_i^a &\longrightarrow &B_i^a.\end{eqnarray}
            By this transformation the CJD Lagrangian transforms in the following way,

           \begin{eqnarray*}\frac{1}{8}\eta (\Omega_{ij}\Omega_{ij}-\frac{1}{2}\Omega_{ii}\Omega_{jj})&\rightarrow& \frac{1}{8}\{\eta(\Omega_{ij}\Omega_{ij}-\frac{1}{2}\Omega_{ii}\Omega_{jj})-\theta\Omega_{ii}-\frac{3}{2\eta}\theta^2\}.\end{eqnarray*}
           Therefore the action becomes,
           \begin{eqnarray}S&=&\frac{1}{8}\int\{\eta(\Omega_{ij}\Omega_{ij}-\frac{1}{2}\Omega_{ii}\Omega_{jj})-\theta\Omega_{ii}-\frac{3}{2\eta}\theta^2\}.\end{eqnarray}
           Einstein introduced cosmological constant to explain static
           nature of the universe. In particle physics, the cosmological
           constant turns out to be a measure of the energy density of the vacuum -
           the state of lowest energy.
            The Einstein-Hilbert Lagrangian for pure gravity
           with cosmological constant is\cite{Pel93}\begin{eqnarray}\mathcal{L_{EH}} & = &
           e(e_I^\alpha e_J^\beta R_{\alpha\beta}^{IJ}(\omega(e))+2\lambda)\end{eqnarray}
           where $e_I^\alpha$ is the tetrad field,
            $R_{\alpha\beta}^{IJ}(\omega(e))$ is the curvature of the unique
           torsion-free spin-connection $\omega^\alpha_{IJ}$, compatible with
           $e_I^\alpha$ and e = $\frac{1}{det(e_I^\alpha)}$. The \textquotedblleft CP-violating"
           $\theta$ term in the action(25) does not contribute to the equations of
           motion and to the energy momentum tensor. $ \mathcal{P} \equiv
           -\frac{1}{16\pi^2}tr^*F^{\mu\nu}F_{\mu\nu}$ is called the
           \textquotedblleft Pontryagin density" and $q = \int\mathcal{P}d^{4}x$, being a
            topological invariant, is  called the Pontryagin index.
            The presence of the $\theta$-term in the action does not alter the theory
            perturbatively, since it is a topological term. Therefore,  keeping aside the topological
           consideration of the $\theta$-term, we can compare the other terms of the
             action(25)   with that of the Einstein-Hilbert   Lagrangian(26) and then we can write
                 $\lambda = -\frac{3}{32}\theta^2$.    At present, in the large scale
                  structure of spacetime, $\lambda$ is found to be zero.   Thus one
                  may assume $\theta \approx 0$ in macroscopic gravity in vacua.
                   But this may  not be the case in micro-domain of matter or in early
                   universe. This suggests that in Planck scale, the torsion and
                   hence the cosmological term may have a significant role.
            \vspace{4 mm}

            \noindent\textbf{Acknowledgement}

              \vspace{1 mm}
           I wish to thank Prof. Pratul Bandyopadhyay, Indian Statistical Institute, for his remarks and suggestions on this
           problem.
                 \vspace{4 mm}

            \noindent\textbf{References}

             \bibliographystyle{unsrt}
                \bibliography{tbib}
\end{document}